\documentclass[prb,showpacs,showkeys,twocolumn,superscriptaddress,10pt,floatfix]{revtex4-1}

\usepackage{graphicx}
\DeclareGraphicsExtensions{.pdf,.png,.jpg}
\usepackage{epsfig}
\usepackage{transparent}
\usepackage{natbib}
\usepackage[colorlinks=true,linkcolor=blue]{hyperref}
\usepackage{breakurl}
\usepackage{color}
\usepackage[dvipsnames]{xcolor}
\usepackage{soul}
\usepackage[normalem]{ulem} 
\usepackage[toc,page]{appendix}

\usepackage{chngcntr}
\usepackage{float}
\usepackage{gensymb}

\begin{document}

\title{Collective magnetization dynamics in nano-arrays of thin FePd discs}

\author{Agne Ciuciulkaite}
\email[]{agne.ciuciulkaite@physics.uu.se}
\affiliation{Department of Physics and Astronomy, Uppsala University, Box 516, SE-75120 Uppsala, Sweden}

\author{Erik \"Ostman}
\affiliation{Department of Physics and Astronomy, Uppsala University, Box 516, SE-75120 Uppsala, Sweden}

\author{Rimantas Brucas}
\affiliation{Department of Engineering Sciences, Uppsala University,
Box 534, SE-751 21 Uppsala, Sweden}
\affiliation{\AA ngstr\"om Microstructure Laboratory, Uppsala University,
Box 534, SE-751 21 Uppsala, Sweden}
    
\author{Ankit Kumar}
\affiliation{Department of Engineering Sciences, Uppsala University,
Box 534, SE-751 21 Uppsala, Sweden}

\author{Marc A. Verschuuren}
\affiliation{Philips Research Laboratories, High Tech Campus 4, Eindhoven, The Netherlands}

\author{Peter Svedlindh}
\affiliation{Department of Engineering Sciences, Uppsala University,
Box 534, SE-751 21 Uppsala, Sweden} 

\author{Bj\"orgvin Hj\"orvarsson}
\affiliation{Department of Physics and Astronomy, Uppsala University, Box 516, SE-75120 Uppsala, Sweden}

\author{Vassilios Kapaklis}
\email[]{vassilios.kapaklis@physics.uu.se}
\affiliation{Department of Physics and Astronomy, Uppsala University, Box 516, SE-75120 Uppsala, Sweden}

\date{\today}

\begin{abstract}
We report on the magnetization dynamics of a square array of mesoscopic discs, fabricated from an iron palladium alloy film. The dynamics properties were explored using ferromagnetic resonance measurements and micromagnetic simulations. The obtained spectra exhibit features resulting from the interactions between the discs, with a clear dependence on both temperature and the direction of the externally applied field. We demonstrate a qualitative agreement between the measured and calculated spectra. Furthermore, we calculated the mode profiles of the standing spin waves excited during a time-dependent magnetic field excitations. The resulting maps confirm that the features appearing in the ferromagnetic resonance absorption spectra originate from the temperature and directional dependent inter-disc interactions. 

\end{abstract}

\maketitle

\section{\label{Introduction}Introduction}
\begin{figure}
\includegraphics[width=0.75\columnwidth]{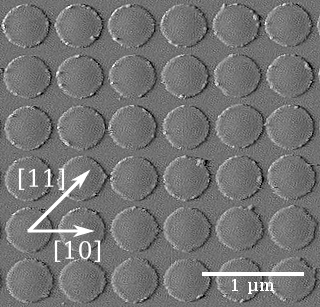}
\caption{Atomic force microscopy image of a square array of Fe$_{20}$Pd$_{80}$ alloy discs with the high symmetry directions indicated as [10] and [11] by arrows. 
\label{fig1} 
}
\end{figure}

Arrays of closely packed mesoscopic magnets provide a rich playground for investigations of collective magnetization dynamics. The (stray field induced) interaction between the discs forms a link between the internal magnetization dynamics of the elements and the global response of the system\cite{Krawczyk_MagnonicCrystals,Nisoli_2017_NPHYS_comment,Bhat_PRB_2018,Jungfleisch_PRB_2016}. Magnetic discs are interesting in this context, due to the richness of internal magnetic textures and the absence of shape induced anisotropy in their plane. Discs of certain radius and height ratio exhibit a ground state referred to as a \textit{vortex}\cite{Shinjo930, PhysRevLett.83.1042} characterized by an in-plane magnetic flux closure. Since the magnetic moments are curling in-plane of the discs, the stray field from the discs is negligible when vortices are formed, in stark contrast to the \textit{collinear state} \cite{PerpendicularFerromagneticResonanceInSoftCylindricalElements, PhysRevB.92.054420, srep25196, Erik_HysteresisFreeSwitching}.
The application of an external magnetic field drives the vortex core out of the center, towards the edge of the disc. At a given field, the vortex is annihilated and the magnetic moment is aligned parallel to the direction of applied field (collinear state). 
When the discs are in a collinear state, their stray fields result in inter-disc interactions. 
Previous investigations of iron-palladium (Fe$_{20}$Pd$_{80}$) discs, arranged in a square array (See Figure \ref{fig1}) showed that a change of temperature is sufficient to alter the magnetization dynamics\cite{Erik_HysteresisFreeSwitching}. Above a given temperature the energy barrier for switching from a vortex to the collinear state and vice versa was even found to be free from hysteresis. As a consequence the magnetization state of the system under the certain applied field becomes bi-stable: the vortex and the collinear state have the same energy.

Here we explore the effect of changes in the inter-disc interactions on the magnetization dynamics of soft magnetic iron-palladium alloy discs, using ferromagnetic resonance (FMR) and micro-magnetic simulations\cite{Dutra_PRB_2019}.  The changes in the inter-disc interactions are obtained by rotating the magnetisation of the discs as well as altering the stray field from the discs by changing the sample temperature. 
\section{\label{sec:Methods}Methods}
 
\subsection{\label{Sample_description}Sample}

The investigated array consists of iron-palladium alloy (Fe$_{20}$Pd$_{80}$) discs arranged in a square lattice as illustrated in Figure \ref{fig1}.  
Each disc has a radius of 225 nm and a thickness of 10 nm, with a center-to-center distance of the discs equal to 513 nm in the [10] direction. This results in an inter-disk distance of 53 nm and 275 nm along the [10] and along the [11] direction, respectively. A detailed description of the sample preparation is provided by \citet{Erik_HysteresisFreeSwitching}.

\subsection{\label{subsec:FMR_measurements}Ferromagnetic resonance measurements}

Magnetization dynamics of the Fe$_{20}$Pd$_{80}$ disc arrays were measured using X-band cavity FMR equipped with with variable-temperature sample holder. A static magnetic field was applied in-plane, while a time-dependent spatially uniform (wave vector $\vec{k}$=0) magnetic field excitation, with a frequency of 9.8 GHz was applied perpendicular to the plane of the sample \cite{PerpendicularFerromagneticResonanceInSoftCylindricalElements,CollectiveModesInMagnonicVortexCrystals}. The static magnetic field was swept from 0 to 300 mT and the measurements were carried out at temperatures ranging from 80 K to 293 K. The strength of the applied static field ensured that discs were measured in the collinear state.

A second set of FMR measurements were performed using a vector-network-analyzer (VNA) at a room temperature, by an in-plane field sweep, utilizing a coplanar waveguide. The detailed description of this measurement setup is described by \citet{staticanddynamicmagneticpropertiesRimantas}. The linewidth $\Delta H_t$ versus frequency $f$ data obtained from these measurements were fitted to a linear function, extracting the Gilbert damping coefficient $\alpha$. The extracted value for $\alpha$ was 1.8$\times$10$^{-2}$ and was later employed as a parameter in the micromagnetic simulations of the magnetization dynamics in the Fe$_{20}$Pd$_{80}$ disc array. It was adjusted in order to match the calculated FMR absorption peak amplitude to one measured experimentally.

\subsection{\label{sec:Micromagnetic_simulations}Micromagnetic simulations and standing spin wave map calculations}

Micromagnetic simulations were performed using M{\scriptsize U}M{\scriptsize AX}3\cite{mumax3_verification}. The exchange stiffness constant, $A_{ex}$, defining the inter-spin coupling in the magnetic material was adjusted to qualitatively reproduce the experimental observations. It is known that $A_{ex}$ is temperature dependent and decreases with increasing temperature \cite{TempDependenceofAexMulazzi}. Therefore a higher temperature implies softer standing spin wave modes, excited within the discs. 
After the initial simulations, a value of $A_{ex}$=3.36 pJ/m was used for the subsequent micromagnetic simulations, in order to qualitatively reproduce the experimentally observed FMR spectra features. A broader discussion and motivation regarding this choice is provided by \citet{MasterThesis}. Finally, the Gilbert damping parameter, $\alpha$, describing the losses in the system and proportional to the FMR absorption linewidth was chosen to be 1.9$\times$10$^{-2}$, after initially taking the value determined from the VNA-FMR measurements.

In the micromagnetic simulations, four discs with 225 nm radius and 10 nm thickness were placed in 2-by-2 square lattice with a lattice parameter of 513 nm. The in-plane cell size was defined as 0.513(18)$\cdot l_{ex}$, where $l_{ex}$ is the exchange length, a material parameter determined by the \textit{$A_{ex}$} and \textit{$M_{sat}$} \cite{mumax3_verification}. The cell size along the $z$-direction, i.e. the thickness of the structure, was set to 5 nm for all simulations. Periodic boundary conditions (PBC)\cite{mumax3_verification} were applied in both lateral directions (PBC$_x$=PBC$_y$=3, PBC$_z$=0). The FMR simulations were performed applying a static magnetic field in-plane of the lattice, relaxing the system, and then applying a time-dependent field excitation out-of-plane, with an analytical expression of $A\cdot$sin(2$\pi ft$) or $A\cdot$sinc(2$\pi ft$). The static magnetic field is applied at an angular offset of 2\degree from the principal in-plane directions, in order to lift any degeneracy in the simulations, related to the high symmetry directions we will be investigating, being parallel to the [10] and [11] direction of the disc lattice. 
The amplitude of the time-dependent excitation was $A$=5 mT, the frequency was $f$=9.8 GHz, the duration of the sinusoidal time-dependent magnetic field excitation was 10 ns, and the sampling period was 1 ps. The frequency bandwidth for the \textit{sinc} function excitation was 20 GHz. The temperature-dependent saturation magnetization, $M_{sat}(T)$, was used to account for the temperature dependence of the FMR response of the soft magnetic discs. The Fe$_{20}$Pd$_{80}$ alloy has a Curie temperature, $T_C$, of 463 K\cite{Erik_HysteresisFreeSwitching}. To a first approximation the temperature dependence of the magnetization can be described by the modified Bloch behaviour:

\begin{equation}
M_{sat}(T)=M_{sat}(0)\bigg[1-\bigg(\frac{T}{T_C}\bigg)^\beta\bigg],
\end{equation}
where $M_{sat}(0)$ is the saturation magnetization at 0 K and is 5.9$\times$10$^5$ A/m for the Fe$_{20}$Pd$_{80}$ alloy and $\beta$=1.69\cite{Erik_HysteresisFreeSwitching}. Here we would like to note that we do not account for any other temperature induced effects in our simulations than the temperature dependence of the saturation magnetization.

A complementary method for describing the observed features in the FMR spectra are standing spin wave (SSW) mode maps. The maps were calculated from the spatial micromagnetic evolution of the magnetization as excited by the time-dependent magnetic field. A fast Fourier transform (FFT) for each of the discrete spatial elements was performed, resulting in the spatial maps of the amplitude of the magnetization dynamics.

\section{\label{sec:Results}Results and Discussion}

\subsection{\label{subsec:ExpRes}FMR}

\begin{figure}
	\centering
	\includegraphics[width=0.85\columnwidth]{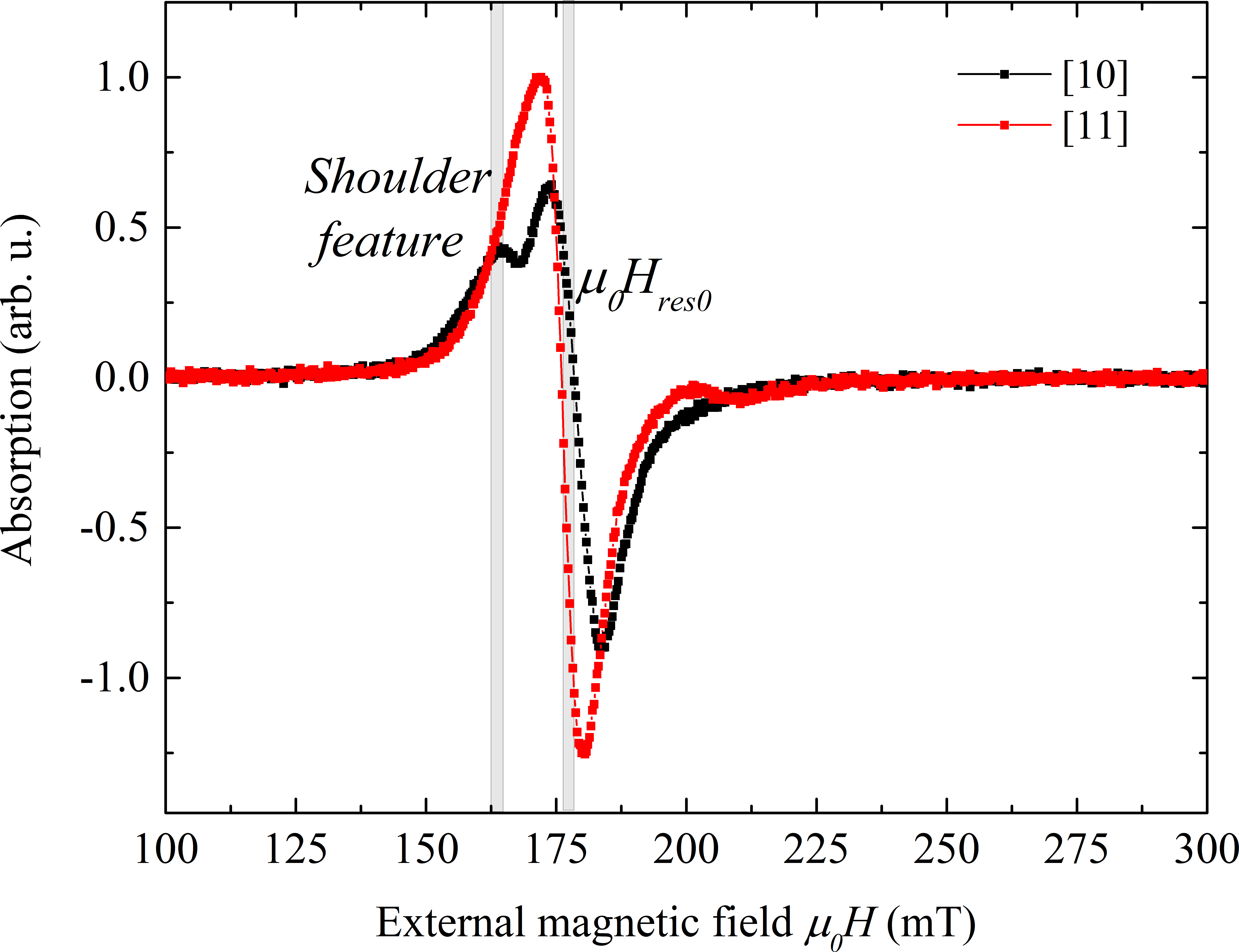} 
	\caption{\label{fig2} The measured FMR spectra of square arrays at room temperature with the field applied along [10] and [11] directions. The shaded grey regions indicate characteristic FMR absorption features of the investigated structure: a \textit{shoulder feature}, appearing before the main absorption peak, and a main FMR field, $\mu_0H_{res0}$.}
\end{figure}

The FMR spectra measured at room temperature, with the static field applied along [10] and [11] in-plane directions of the array, as indicated in Figure \ref{fig1}, are shown in Figure \ref{fig2}. The ferromagnetic resonance shifts to slightly higher applied magnetic fields, when the field is applied along [10] direction as compared to the  [11] direction.

This shift reflects the effect of the inter-disc interaction strength, due to the change in the applied static magnetic field and hence the stray field direction. 
Furthermore, the measured FMR absorption spectra exhibits a split in the main absorption peak in the [10] direction. We will call this feature before the main absorption peak a \textit{shoulder feature} (See Figure \ref{fig2} in the following). The change in interaction strength is due to a modification of inter-disc coupling by changing the magnetisation direction of the discs. Effectively, along the [10] direction a disc has two nearest neighbours while along the [11] direction number of interacting nearest neighbours is doubled as described below.

\subsection{\label{subsec:disc}Micromagnetic simulations on a single disc}
\begin{figure}[htb!]
	\centering
	\includegraphics[width=\columnwidth]{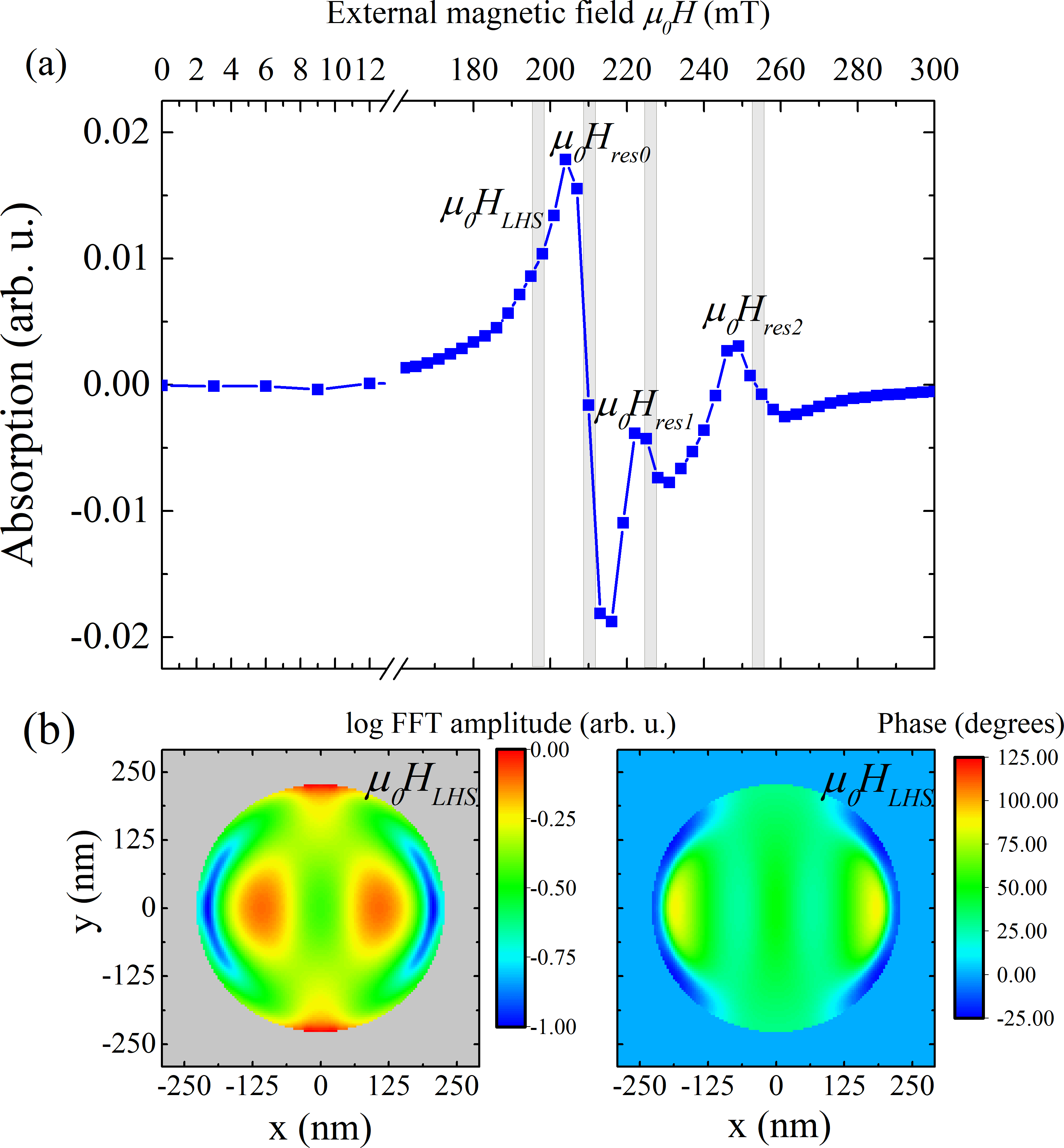} 
	\caption{\label{fig3} (a) FMR absorption spectrum for a single disc, taken as a linecut from the map in Figure \ref{fig_A1} (see Appendix \ref{AppendixdiscMap}) at 9.8 GHz frequency. The shaded grey regions indicate external magnetic fields at which the SSW profiles ($m_z$ component, amplitude of precession) were calculated; (b) FFT amplitude and phase maps calculated at a shoulder feature at 197 mT applied magnetic field.} 
\end{figure}

Micromagnetic simulations of a single disc were carried out to identify the basic features of the elements, $\it{i.e.}$ the response of the discs in absence of interactions. A FMR frequency versus applied static field map was calculated for a wide range of frequencies, using the expression $A\cdot$sinc(2$\pi ft$) for the time-dependent magnetic field excitation and is provided in Appendix \ref{AppendixdiscMap}, Figure \ref{fig_A1}. Since the FMR measurements were carried out at the frequency of 9.8 GHz, a line cut along this frequency was taken and is shown as an FMR absorption spectrum in Figure \ref{fig3} (a). The shaded grey regions indicate the strength of the applied static magnetic field, for which the SSW modes excited in the disc were simulated and are presented as maps of normalized magnetization precession amplitude and phase in Figure \ref{fig3}~(b) and in Appendix \ref{AppendixB} Figure \ref{fig_B1}. 

Figure \ref{fig3} (b) presents the SSW maps for an amplitude and phase for applied magnetic field corresponding to the region where the \textit{shoulder feature} in 
Figure \ref{fig2} is observed. It is indicated on the left-hand-side (LHS) of the main absorption peak as $\mu_0H_{LHS}$ in Figure \ref{fig3} (a). This mode is mainly constrained near the middle of the disc and along a line parallel to the direction of the applied magnetic field. In the remainder of this communication we will refer to the area in the middle of a disc as the \textit{central} area. The largest precession takes place at two symmetric points outside of the disc center, along the direction of static magnetic field. In addition to these symmetry points, the magnetic moments also resonate at the edges of the disc along a line perpendicular to the magnetic field direction. In the following, we will refer to this mode as a \textit{perpendicular edge} mode. To SSW modes appearing at the edges along a line parallel to the direction of the applied magnetic field, we will be referring to as \textit{parallel edge} modes.
The SSW spatial mode maps calculated at higher magnetic fields indicate the following spatial mode profiles: at 210 mT field - 
uniform precession, at 226 and 255 mT - edge modes (see Appendix \ref{AppendixB} Figure \ref{fig_B1}).

 \subsection{\label{Res:array_sim} Micromagnetic simulations of square arrays}

\begin{figure}
	\includegraphics[width=1.\columnwidth]{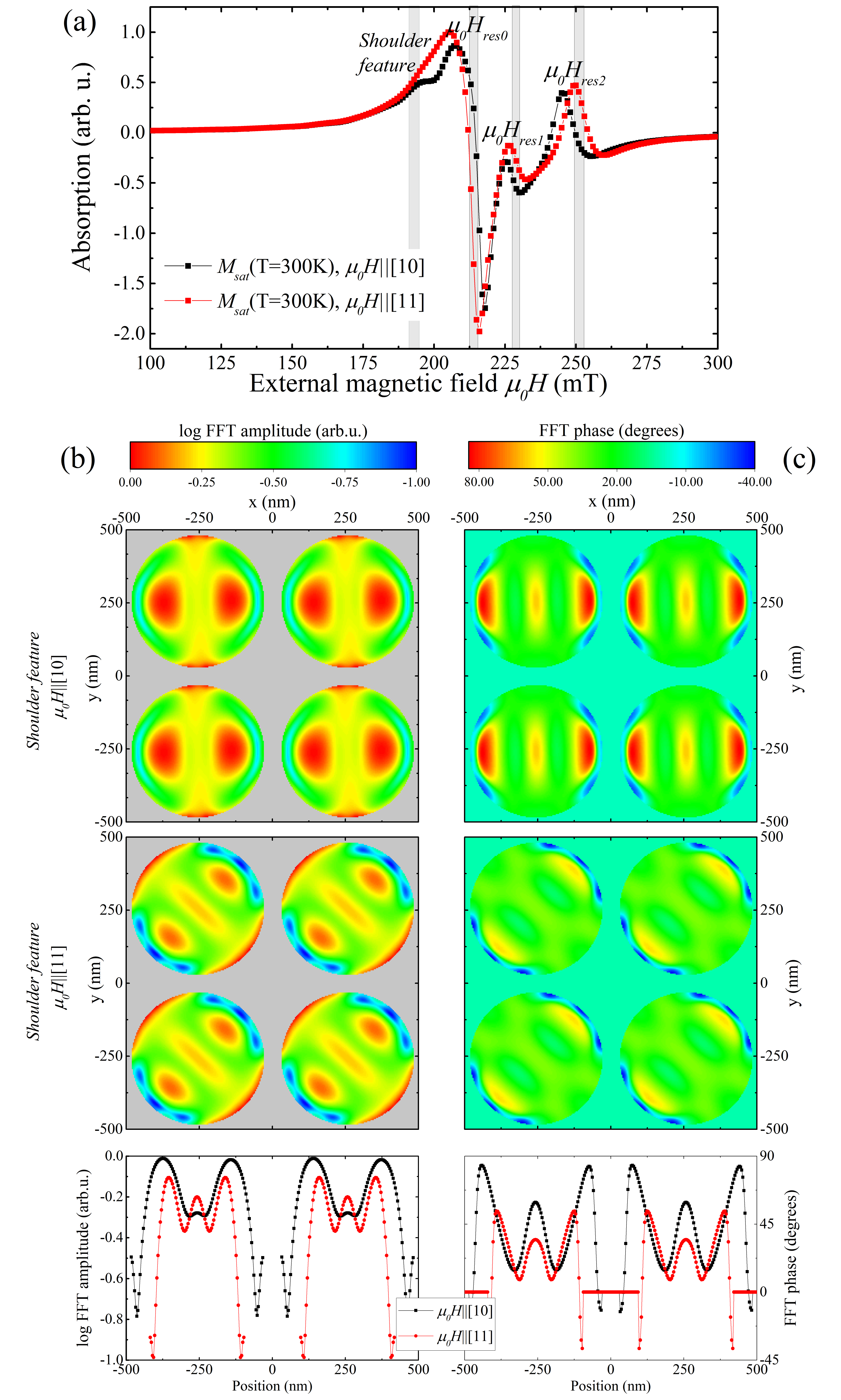}
	\caption{\label{fig4} (a) Computed FMR absorption spectra with  $M_{sat}(T=300 K)$ and static magnetic field applied along the [10] and [11] directions. Grey shaded regions represent external magnetic fields at which SSW modes were calculated; (b) spatial amplitude and (c) phase maps of SSW modes calculated for the two applied field directions. Color bars indicate normalized amplitude and phase angles in degrees; the bottom panels in (b) and (c) represent amplitude and phase linecuts in the respective maps along the [10] and [11] directions.} 
\end{figure}

Subsequent simulations were carried out on square arrays containing interacting discs, but otherwise identical to the building block described in the previous section. In order to explain the \textit{shoulder} feature in the FMR spectra for the [10] direction (Figure \ref{fig2}) and to further investigate which SSW modes are excited in the discs, spatial amplitude and phase maps (see Figure \ref{fig4} (b)-(c)) were calculated at external magnetic fields of 191 mT and 186 mT along the [10] and [11] directions, respectively.

The \textit{shoulder} feature on the left-hand-side of the FMR absorption peak, observed in the experimentally measured spectra (Figure \ref{fig2}), is also present in the calculated results, when the external magnetic field is applied along the [10] direction (Figure \ref{fig4} (a)). 
In this case, calculated SSW mode spatial maps show that the magnetic moments are strongly out of phase at the \textit{parallel edge} and \textit{center} areas of the discs. When the direction is changed to [11], the interaction strength between the discs is modified, due to effective increase of nearest neighbors along the [11] direction and the magnetic moment precession amplitude significantly increases in the \textit{center} and \textit{perpendicular edge} areas. Simultaneously, the phase angles in the \textit{center} and at the \textit{parallel edges} of a disc decrease, while the overall phase angle distribution over each disc becomes more uniform as compared to the case when external field is along the [10] direction.  
Furthermore, the amplitude of the \textit{perpendicular edge} mode and the extension over the edges is smaller for the [10] external field direction as compared to the [11] direction. This again is a hallmark of the effect the proximity to the nearest neighboring discs has on the magnetization dynamics. 
Comparison of the SSW mode amplitude and phase spatial maps, to those of an isolated single disc reveals that the amplitude of magnetic moments in the \textit{center} of an isolated disc is smaller as compared to the same area in the arrays.  

These results clearly demonstrate the effect of the stray field coupling between discs in an array on the internal magnetization dynamics of the elements. Modifying the inter-disc interaction strength alters the magnetic moment fluctuations at the center of a disc which results in appearance (or disappearance) of the \textit{shoulder} feature in an FMR absorption spectrum. A further important observation arising from the simulations, is the absorption modes appearing at higher fields after the main absorption peak. 
These features, indicated as $\mu_0H_{res1}$ and $\mu_0H_{res2}$ at around 229 mT and 251 mT, respectively in Figure \ref{fig4} (a)), are not observed in the experimental FMR spectra (Figure \ref{fig2}). They arise from the idealised circular shape of the discs in the simulations. In real samples the lithographically fabricated elements are imperfect, leading to the absence of these modes \cite{Jungfleisch_ASI_AMR}. The origin of spin wave maps for these simulated modes is discussed in more detail in the Appendix \ref{AppendixC}.

\subsection{\label{Results_calculated_FMR_temp}Ferromagnetic resonance response of iron-palladium arrays: comparison of experiments and simulations} 

\begin{figure}
	\includegraphics[width=0.95\columnwidth]{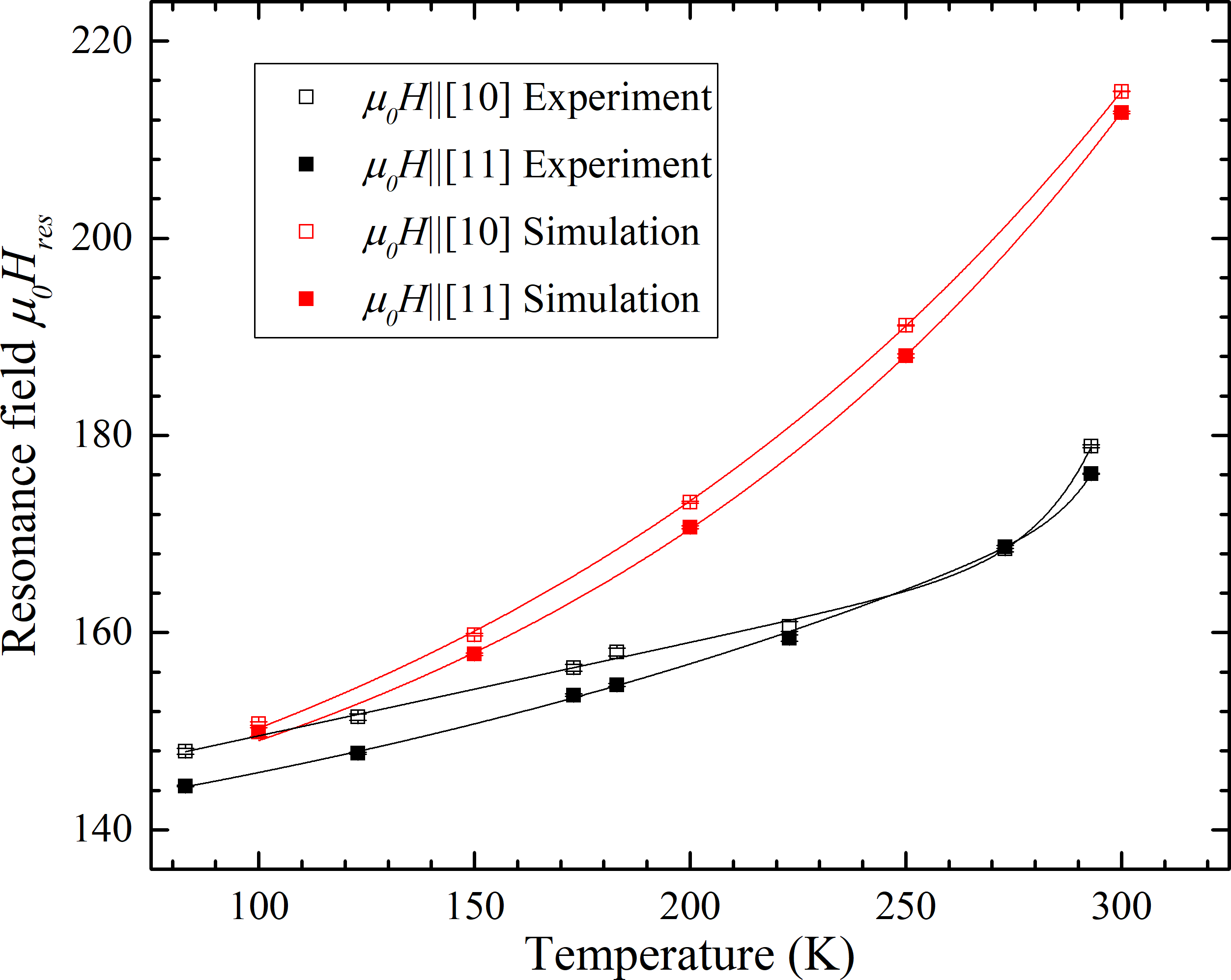}
	\caption{\label{fig5}Comparison of the resonance fields obtained via micromagnetic simulations (red squares) and experimental values (black squares) extracted from FMR measurements  of the Fe$_{20}$Pd$_{80}$ alloy disc array. The external static magnetic field was applied along [10] and [11] directions (empty and filled symbols, respectively). Lines 
	are guides to the eye. Error bars and resonance fields were obtained from fitting of FMR amplitude to a Lorentzian peak profile.} 
\end{figure}

A comparison of the measured and calculated spectra is provided in Figure \ref{fig5}, where we display the change in the ferromagnetic resonance field with increasing temperature, for applied fields in the [10] and [11] directions. The strength of the interaction alters the resonance field as observed in the measured FMR spectra. Same qualitative changes are observed in the micromagnetic simulations. The resonance field increases as the temperature is increased (see Figure \ref{fig5}). 
 Thermal fluctuations become more prominent with increasing temperature and a stronger magnetic field is needed to align the moments, shifting the resonance field to higher values. This trend is seen in both calculated and measured FMR spectra. At low temperatures the experimental and calculated resonance fields are comparable. At elevated temperatures the resonance fields calculated from micromagnetic simulations increase faster than the measured values. The micromagnetic simulations offer a qualitative agreement with the experimental data which is more clearly seen from a comparison of measured (Figure \ref{fig2}) and calculated FMR absorption spectra (Figure \ref{fig4}(a)) at elevated temperatures.

In Figure \ref{fig5} we can further observe that the splitting between the simulated resonance field values along [10] and [11] directions, is always present in the investigated temperature range. The curves for the [10] direction lie higher in field, compared to those for the [11] direction in the investigated temperature range. In related studies, where neighboring elongated micromagnetic elements were antiferromagnetically coupled by stray fields, resonance field shifts depending on the strength of the interaction, were also reported \cite{DEMAND2002228,Guido_permalloyRectangles}.  However, the difference in resonance field for [10] and [11] directions ($H_{res[10]}-H_{res[11]}$) calculated from micromagnetic FMR results (by fitting amplitude to a Lorentzian profile peak and extracting its position), follows a different trend as compared to the resonance field 
extracted from experimentally obtained data (Figure \ref{fig5}). This can be attributed to the fact that while exchange stiffness and Gilbert damping parameters are temperature dependent, in micromagnetic simulations we kept these parameters constant for all temperatures. Furthermore, no thermal fluctuations were accounted for in micromagnetic simulations. 

Finally, we would like to comment upon the scaling of the experimental resonance field difference $H_{res[10]}-H_{res[11]}$ versus temperature. 
The difference decreases monotonically up to some temperature, where it effectively becomes zero, while beyond that it seems to increase again. This behaviour potentially relates to the internal magnetization dynamics within the discs. As already reported previously by \citet{Erik_HysteresisFreeSwitching}, a temperature range exists where bi-stability is attainable for the collinear and vortex states, which should also be accompanied by a strong modification in the magnetization dynamics of the individual discs. This range was found to be independent of applied field direction, for temperatures above $\approx$~220~K from measurement protocols involving very low frequencies (hysteresis curves, recorded employing the magneto-optical Kerr effect and field cycling of 0.4 Hz). In the present study, $H_{res[10]}-H_{res[11]}$ becomes zero at $\approx$~270~K, considerably higher than that in \citet{Erik_HysteresisFreeSwitching}. The time scale of dynamics, probed in this work, is much shorter (sub-ns, FMR measured at 9.8 GHz), hinting for this strong apparent temperature shift. A more thorough survey of this effect would greatly benefit from more detailed micromagnetic simulations, incorporating also the magnetization dynamics of the material at all relevant length-scales (inter- and intra-disc).

\begin{figure}[htb!]
	\includegraphics[width=1\columnwidth]{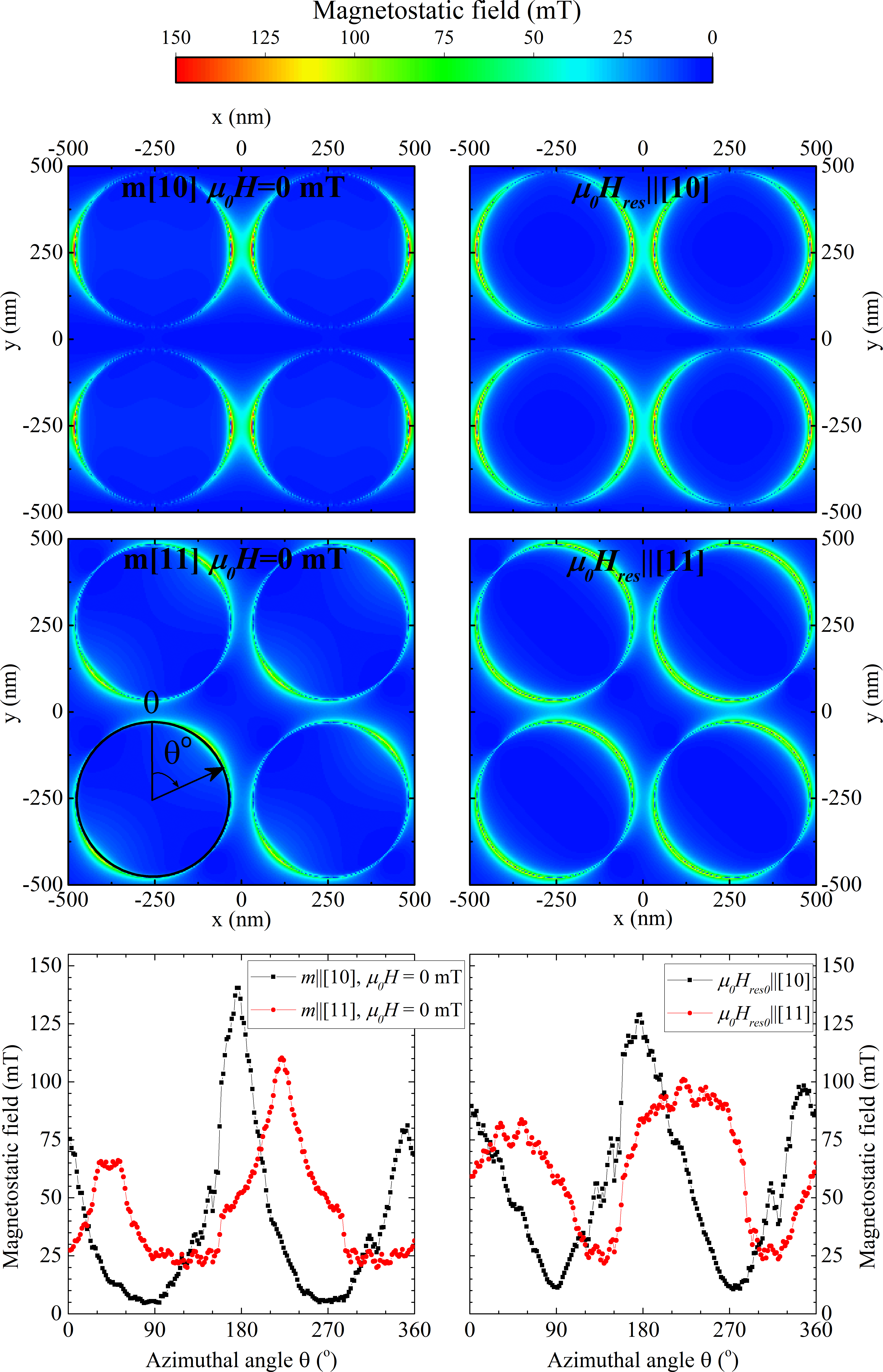}
	\caption{\label{fig6} Spatial distribution maps of demagnetizing field strengths, obtained from micromagnetic simulations of arrays with initial magnetization along [10] and [11] directions (first and second row, respectively) relaxed in a 0 mT static magnetic field and a field corresponding to the ferromagnetic resonance field (left and right columns, respectively). Bottom panels represent polar plots of the demagnetizing field, along the rim of a disc, as indicated by a circle in the ``m[11]$\mu_0H$=~0~mT" map.} 
\end{figure}	

\begin{figure}
	\includegraphics[width=\columnwidth]{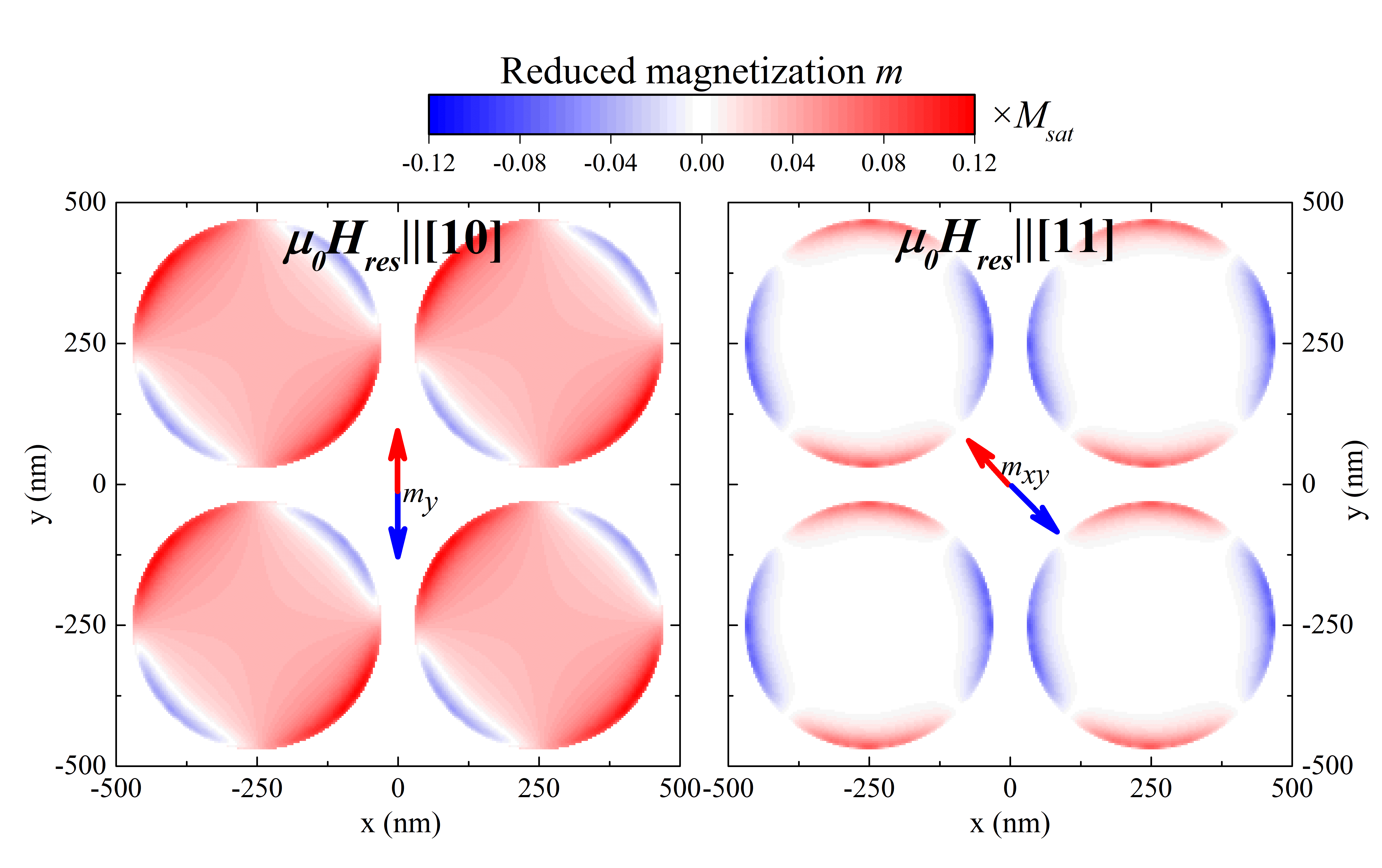}
	\caption{\label{fig10} Spatial maps of magnetization components distribution perpendicular to the direction of the external magnetic field. Maps were obtained from micromagnetic simulations of arrays with initial magnetization along [10] and [11] directions relaxed in a static magnetic field corresponding to $H_{res}$ in magnitude for each of the directions.} 
\end{figure}

The shift in the  resonance field is attributed to a difference in the stray field induced inter-disc coupling, along the [10] and [11] directions. 
In Figure \ref{fig6}, we show the spatial maps of the demagnetizing field amplitude, computed for arrays with initial collinear magnetization states, along [10] and [11] directions. Having the magnetisation of the elements along the [11] direction results in an increased stray field coupling between neighboring discs. 
Clearly, a simple point-dipole-like approximation is not sufficient to describe the effective magnetostatic coupling. Even though the spacing between neighboring discs in the [10] direction is smaller - and thus one would expect a stronger coupling - the demagnetizing field strength is distributed broadly along the disc perimeter for the [11] case (see the bottom panels in Figure \ref{fig6}). This leads to an stronger coupling for the applied fields along the [11] direction. 
The root of these effects are clearly seen in Figure \ref{fig10}, where we provide spatial maps of static magnetization components perpendicular to the direction of applied magnetic field. 
A map of the $m_y$ component with $H_{res}||$[10], shows a dominant positive direction for these, in accordance to the 2\degree offset of $H_{res}$ from the [10] direction in our simulations. 
When $H_{res}||$[11], the magnetization components $m_{xy}$ perpendicular to applied fields, obtain non-zero values only at the discs' rim and with maxima at the positions where the gap to neighboring disc is minimum. In contrast to the case where $H_{res}||$[10], now all of the four gaps related to neighboring discs are active. The maps show that $m_{xy}\perp[11]$ is of the same sign at the disc edges along the [10] and [01] directions which further confirming inter-disc coupling via stray fields.

\section{Conclusions}

We have investigated the magnetization dynamics in a square array of Fe$_{20}$Pd$_{80}$ alloy discs by means of FMR measurements and micromagnetic simulations.  
The ferromagnetic resonance is found to increase in field with increasing temperature which we attribute to the decrease of the magnetic moment. The results were qualitatively reproduced and confirmed by micromagnetic simulations. 
The effects of stray field induced interaction of the islands is seen in the orientation dependence of the resonance field as well as the fine structures of the resonance. The inter-island interaction is found to be larger when the magnetisation of the islands are in the [11] as compared to [10] principal directions. The origin of this effect arises from the transverse component of the magnetisation, which enhances the inter-island interactions in the transverse directions, illustrating the need of including inner magnetic structure of the magnetisation to explain both dynamic and static results.   

The results presented in this study, show that micromagnetic simulations are suitable for investigations of spatial profiles of SSW modes, excited in arrays of stray field coupled nanomagnets. These can be utilized for analyzing and designing the microwave response of extended arrays of thermally active and interacting nanomagnets \cite{Farhan_2013_NPHYS, Kapaklis_2014_NNANO, Andersson_SciRep_2016}, having interesting topological and interaction schemes, such as artificial spin ices\cite{Gliga_2013_PRL, Krawczyk_MagnonicCrystals, Jungfleisch_PRB_2016, Iacocca_2016_PRB, ADFM:ADFM201505165,Ostman_2018_NatPhys,Bhat_PRB_2018}. The latter could enable the design and fabrication of reconfigurable magnonic devices\cite{Iacocca_2016_PRB,Bhat_PRB_2018}.

\begin{acknowledgments}
 The authors would like to acknowledge financial support from the Swedish Research Council (VR), the Swedish Foundation for International Cooperation in Research and Higher Education (STINT) and the Knut and Alice Wallenberg Foundation project ``{\it Harnessing light and spins through plasmons at the nanoscale}'' (2015.0060). This work is part of a project which has received funding from the European Union's Horizon 2020 research and innovation programme under grant agreement no. 737093.
 
\end{acknowledgments}
  
\bibliographystyle{apsrev4-1}  
%

\newpage

\appendix
\counterwithin{figure}{section}
\section{{\label{AppendixdiscMap}}Dispersion relation of a single FePd disc}

\begin{figure}[ht!]
	\centering
	\includegraphics[width=0.9\columnwidth]{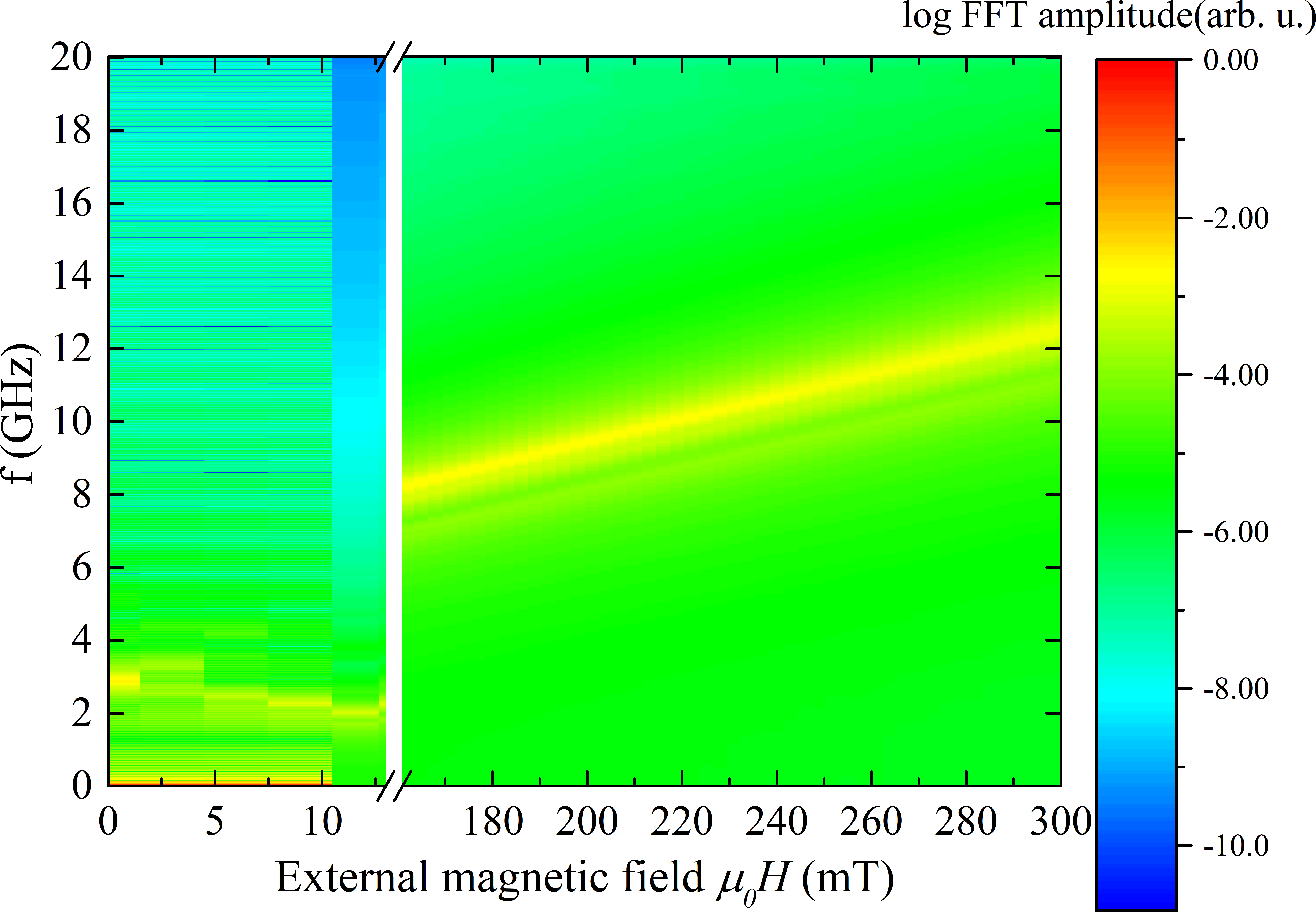}
	\caption{\label{fig_A1} The calculated map of the log and normalized FMR amplitude response to a 20 GHz bandwidth sinc function form magnetic field excitation, for different external static fields.} 
\end{figure}

The complete map of available resonance modes for a single nanodisc resonator is shown in Figure \ref{fig_A1} as a frequency vs external magnetic field map. The color bar represents the log FFT amplitude of spin precession. Below the annihilation field, the spectrum is dominated by low frequency modes, corresponding mostly to the gyrotropic motion of the vortex core\cite{Stroboscopic_PEEM}. 
The highest intensity peak corresponding to a vortex mode shows similar dispersion relation as that for isolated cylinders exhibiting single domain state\cite{DEMAND2002228}. At stronger external magnetic fields above the saturation field, when the disc is in a collinear magnetic state, a significant increase of the resonance frequency is observed and  highest intensity peak follows Kittel-like behaviour of resonance frequency for continuous ferromagnetic films. Below the main resonance peak are two smaller intensity features corresponding to the edge modes.

\section{{\label{AppendixB}}Magnetization maps of single disc SSW modes}

\begin{figure}
	\centering
	\includegraphics[width=0.9\columnwidth]{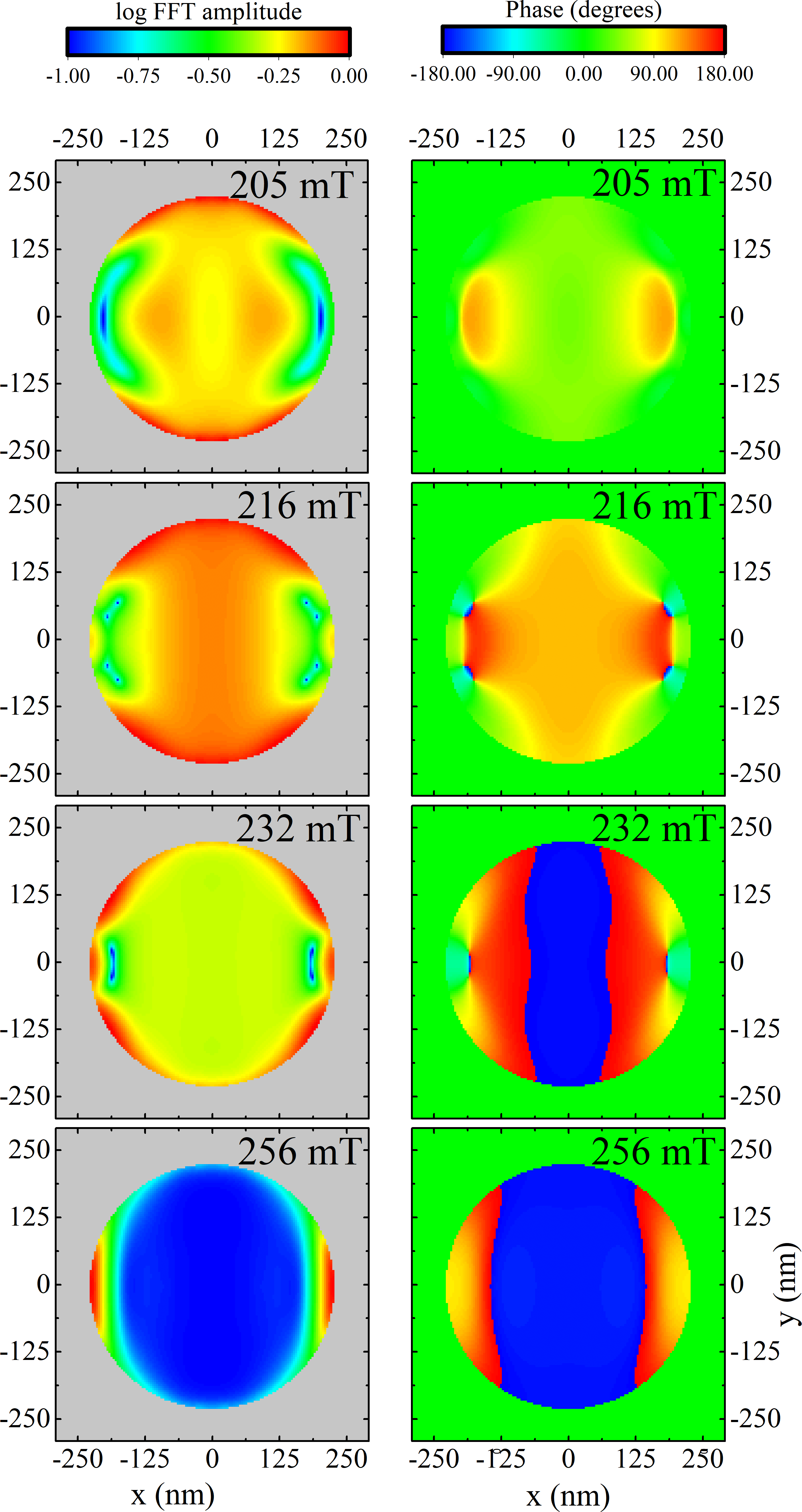}
	\caption{\label{fig_B1} The calculated spatial log and normalized FFT amplitude and phase maps of an isolated single disc at different fields indicated in Figure \ref{fig3} (a). 
	} 
\end{figure}

SSW maps calculated at the resonance field, which is 210 mT for a single disc, reveal that the area of constrained magnetic moments becomes more concentrated at the edges along the direction of an applied magnetic field. Meanwhile, magnetic moment fluctuations become more intense at the \textit{perpendicular edges}. In contrast, the SSW modes calculated at 226 mT and 255 mT external magnetic fields appear at the \textit{parallel edges}. The spatial log normalized FFT amplitude and phase maps calculated at fields just below the ferromagnetic resonance field and above it are shown in Figure \ref{fig_B1}. The evolution of \textit{parallel edge} modes can be observed. Just before and after the ferromagnetic resonance (at 205 mT and 216 mT fields, respectively), spin magnetic moment fluctuation amplitude is uniform throughout the whole area of a disc except for the \textit{parallel edges}. At the additional absorption features observed after the FMR at 232 mT and 256 mT fields, respectively, fluctuation amplitude becomes larger at the edges while throughout the rest of a disc, it goes to zero, in other words, the rest of the spins ``freeze in".  

\section{{\label{AppendixC}}Magnetization maps of edge modes in an array}

\begin{figure}
	\centering
	\includegraphics[width=\columnwidth]{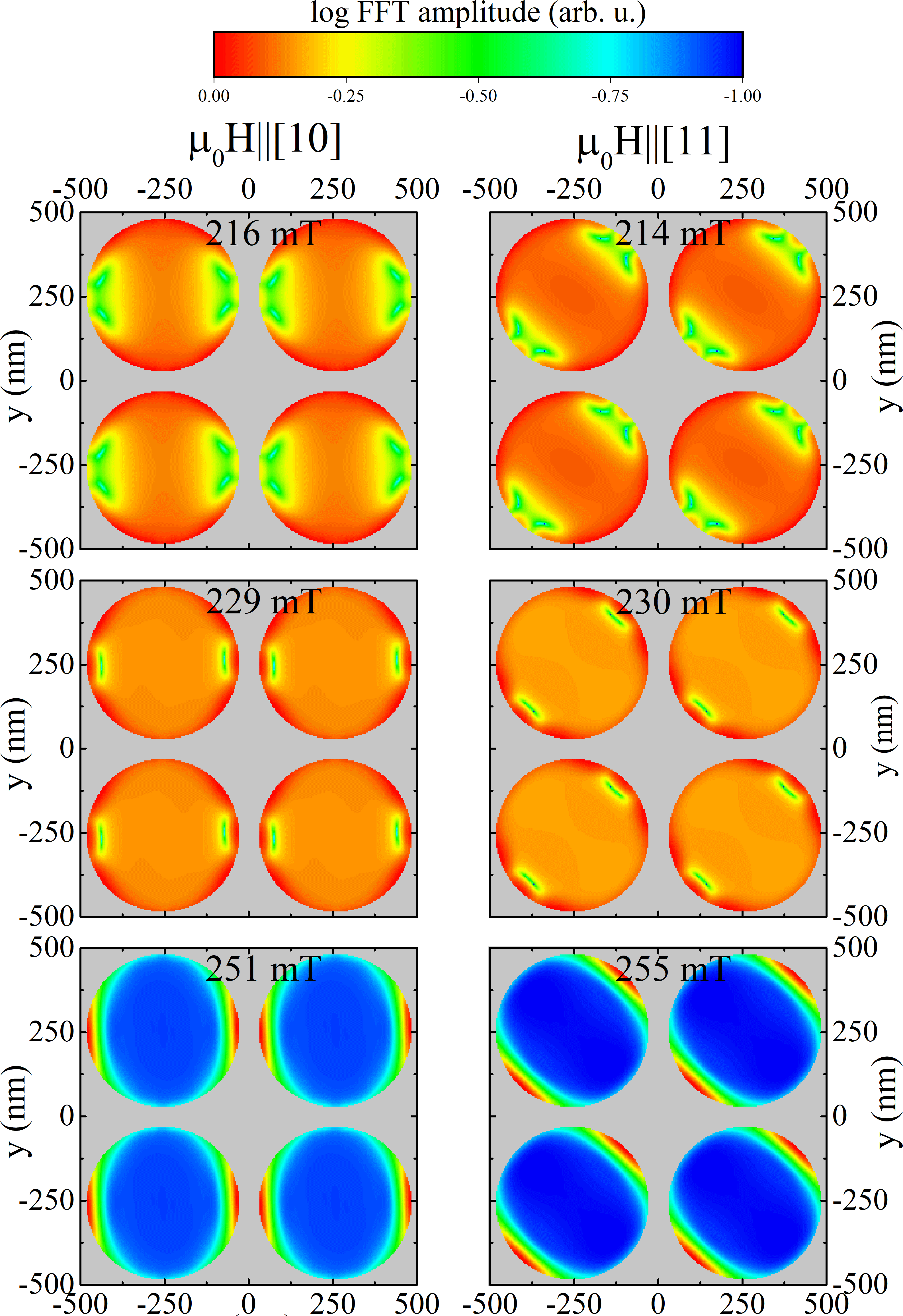}
	\caption{\label{fig_6} Spatial maps of SSW modes calculated for the Fe$_{20}$Pd$_{80}$ alloy disc array at the field values indicated by shaded grey regions in Figure \ref{fig4} (a).} 
\end{figure}

In this section we discuss the resonance and edge modes observed in the calculated FMR spectra in Figure \ref{fig4} (a). At the resonance field, indicated by an arrow at 216 mT field in Figure \ref{fig4} (a), the log FFT magnetisation amplitude of spin magnetic moments is highest at the \textit{perpendicular edge} area as in the single disc case but with the constrained magnetic moment area reduced (compare Figure \ref{fig_B1} amplitude maps of modes at 205 and 216 mT fields with a top panel in Figure \ref{fig_6}.  
When the external magnetic fields are approximately 229 mT and 250 mT, the \textit{parallel edge modes} are excited. Mode maps reveal that in this case the magnetic moments fluctuate the most at the edges along disc coupling direction \ref{fig_6}. This hints that disc interaction occurs through the fluctuating magnetic moments. As a result, along [10] direction the system becomes less stiff and resonance occurs at lower fields when the field is applied along [10] than along [11] direction as can be seen in the absorption spectra in Fig. \ref{fig4} (a) at around 229 and 251 mT fields.

In the real arrays such modes are absent, due to shape imperfections and edge roughness of the discs arising from a lithographic fabrication method. Furthermore, due to computational limitations we calculated significantly smaller amount of nanodiscs than the real sample actually contains \cite{Jungfleisch_ASI_AMR}. This is also further supported by micromagnetic simulations performed accounting for shape and edge imperfections which do not reproduce these modes\cite{MasterThesis}.

\end{document}